\documentclass[aps,prb,amsmath,amssymb,authoryear,numbers,sort&compress, groupedaddress]{revtex4-2}
\usepackage{tikz}
\usepackage{graphicx}
\usepackage{geometry}
\usepackage{color}
\usepackage{setspace}

\begin{document}


\title{Nanoporous Structure of Sintered Metal Powder Heat Exchanger in Dilution Refrigeration: A Numerical Study}


\author{Xiaomin Wu$^{1}$, Yi Liao$^{2}$, Jinxin Zhong$^{3}$, Qing Xi$^{4}$, Lifa Zhang $^{1}$, Jun Zhou 
$^{1}${}*}
\affiliation{1 Phonon Engineering Research Center of Jiangsu Province, Center for Quantum Transport and Thermal Energy Science, Institute of Physics and Interdisciplinary Science, School of Physics and Technology, Nanjing Normal University, Nanjing 210023, China\\
2 Institute for Quantum Science and Engineering, Southern University of Science and Technology, Shenzhen 518055, China\\
3 School of Physics Science and Engineering, Tongji University, Shanghai 200092, China\\
4 Research Center for Advanced Science and Technology, University of Tokyo, 4-6-1 Komaba, Meguro-ku, Tokyo 153-8505, Japan}


\date{\today}

\begin{abstract}
We use LAMMPS to randomly pack hard spheres to simulate the heat exchanger, where the hard spheres represent sintered metal particles in the heat exchanger. We simulated the heat exchanger under different sphere radii and different packing fractions of the metal particle and researched pore space. To improve the performance of the heat exchanger, we adopted this simulation method to explore when the packing fraction is 65\%, the optimal sintering particle radius in the heat exchanger is 30$\sim$35$nm$.
\end{abstract}


\maketitle


\section{Introduction}
Cutting-edge scientific research needs to maintain in an ultra-low-temperature environment, for example, superconducting quantum devices.  $^3$He$/$$^4$He dilution refrigerator can realize this temperature environment. The key to breaking through the extremely ultra-low-temperature is to reduce the Kapitza resistance between $^3$He$/$$^4$He mixture and solid. According to AMM, Kapitza resistance sharply increases with decreasing temperature\cite{pollack1969kapitza, mazo1955theoretical}. The sintered heat exchanger(referred to as porous media) formed by sintred metal powder has large specific surface areas that can reduce the Kapitza resistance. The sintered heat exchanger is composed of packing objects and pore space. In order to increase the effective contact area to increase the heat exchange, the pore space formed by sintered metal powder must be considered to ensure the smooth passage of liquid $^3$He$/$$^4$He mixture. Some works have measured and calculated the size of pore space \cite{nakagawa2019gas,robertson1983properties}, but it is inadequate to guide the sintering process. Therefore, theoretical research on the performance of heat exchangers is also important \cite{nakagawa2019gas,lu2020effects}. The particles formed by sintering usually are regarded as spherical particles of equal size \cite{nakagawa2019gas,robertson1983properties,nakayama1989kapitza, lu2020effects}. And the model of random packing of spheres has been generalized and applied to porous media in many studies\cite{gao2012two, vanderlaan2014he, wang2022temperature, markicevic2019properties, stoyan2011statistical}.

In this work, we randomly packed hard spheres with the same radius in finite space to simulate heat exchangers and numerical calculations on pore space are discussed and the optimized production techniques of heat exchangers are presented.

\section{Simulation calculation of heat exchangers}

\subsection{Simulation calculation methed}
\begin{figure}[h]
\centering
\includegraphics[width=12.5cm]{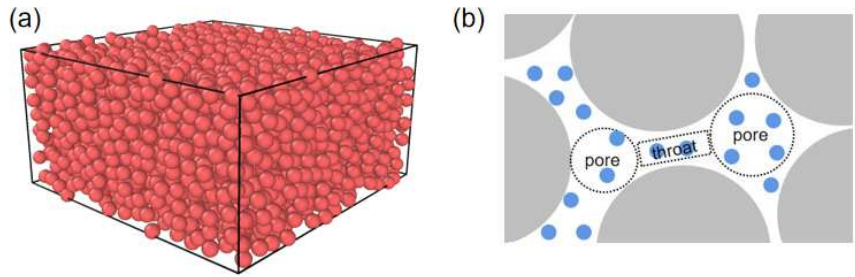}
\caption{\label{fig1}(a) Typical simulation diagram of the sintered particles with radius and packing fraction simulated by the LAMMPS. (b) In the two-dimensional diagram after the simulation of porous media, the gray circle represents the simulated sintered metal particles, the bule circle is the $^3$He/$^4$He atom and the blank part is the pore space, which is composed of pores and throats. Here, we regard the throats as cylinders and pores as spheres.}
\end{figure}

This work used LAMMPS with Morse potential is used to obtain the heat exchanger.  Fig. \ref{fig1}(a) shows the simulation diagram with $R_s$=150 $nm$ and packing fraction of 48$\%$. We regard the pore space as composed of interconnected pores and throats in a similar way to Ref.\cite{gao2012two}. Throats as cylinders and the pores as spheres, as shown in Fig. \ref{fig1}(b) in a two-dimensional diagram. The distance between a sphere and its nearest neighbor sphere was noted as the throat size. The pore size was used to randomly generate pores center outside spheres, which is expressed by the diameter obtained by just contacting the nearest neighbor sphere. The process of sinter formation simulation and pore space numerical calculation is as follows:

\noindent1. Set initial conditions in LAMMPS, input the size of packing space, the diameter and number of spheres, potential parameters, etc.

\noindent2. Output the spherical center coordinates of each packing spheres. 

\noindent3. Import the file of the output spherical center coordinates to OVITO, input the size of the packing sphere, and visualize the simulated the heat exchanger, as shown in Fig. \ref{fig1}(a). 

\noindent4. Import the file of the output spherical center coordinates to MATLAB, Randomly generate N (N cannot be too small) pores center with MATLAB and output pores center coordinates.

\noindent5. Import the file of the output spherical center coordinates and pores center coordinates to MATLAB, Using MATLAB to statistically process spherical center coordinates and pores center coordinates to obtain the average pore size. 

\noindent6. Import the file of the output spherical center coordinates to MATLAB, using MATLAB to statistically process the spherical center coordinates to obtain the average throat size.

\noindent7. Adjust the parameters according to the simulation requirements, and repeat the above steps.

\subsection{Simulation calculation results}
\begin{figure}[h]
\centering
\includegraphics[width=12.5cm]{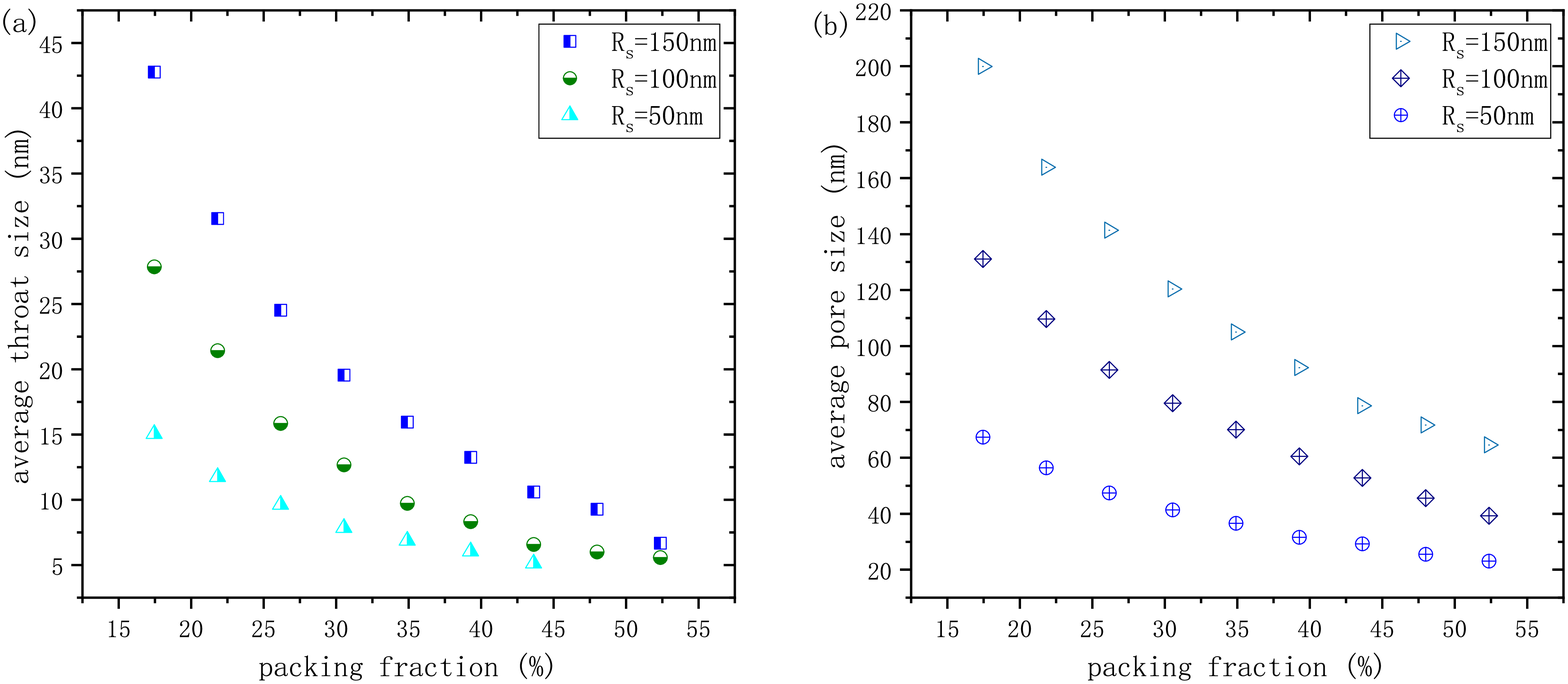}
\caption{(a) Average throat size and (b) average pore size as a function of packing fraction. The radii of simulated sintered metal particles are taken to be 50, 100, and 150 $nm$, respectively.}
\label{fig3}
\end{figure}

One can easily calculate average throat size $d$ and average pore size $D_{h}$ at different packing fractions and different sphere radii. Figs. \ref{fig3}(a) and (b) show the changes in average throat size and average pore size, respectively, with packing fraction $p$ when the radius of the sphere $R_s$=50, 100, and 150 $nm$. It can be seen that with the increase in packing fraction, the change of average throat size and average pore size is smaller, and it tends to be flat under a high packing fraction roughly above 45$\%$.

In addition, according to the hard sphere model, one can have the expression for the specific surface area $\sigma$ as
 \begin{equation}
\sigma=\frac{3}{\rho_m R_s},\label{equ88}
 \end{equation}
where $\rho_m$ is the mass densities of metal particles. Eq. (\ref{equ88}) shows that a smaller radius of the nanoparticle leads to a larger surface area. To verify the feasibility of our simulation calculation, we simulated and calculated results under different packing fractions and different sphere radii of copper particles in Table \ref{tab1}.
According to the obtained parameters, 
the ratio of throat size to pore size is 0.1$\sim$0.2, which agrees well with the observed 0.1 results of sinter in the experiment.

\begin{table}[h]
\setlength{\abovecaptionskip}{0.1cm}
\setlength{\belowcaptionskip}{0.2cm}
\centering
\setlength\tabcolsep{0.7cm}
\caption{\centering{Related parameters obtained by simulations.}}
\begin{tabular}{ccccc}
\hline
packing & specific & particle & pore & throat\\
fraction & surface area & diameter & diameter & size\\
p($\%$) & $\sigma$($m ^2$/g) & D($nm$) & $D_h$($nm$) & d($nm$)\\
\hline
33 & 0.88 & 760 & 320 & 46\\

35 & 1.12 & 600 & 220 & 41\\

40 & 1.20 & 560 & 180 & 27\\

48 & 0.79 & 840 & 240 & 34\\

41 & 1.01 & 660 & 210 & 32\\

42 & 0.86 & 780 & 250 & 45\\

42 & 1.05 & 640 & 200 & 31\\
\hline
\end{tabular}
\label{tab1}
\end{table}

\begin{figure}[h]
  \centering
  \includegraphics[width=9cm]{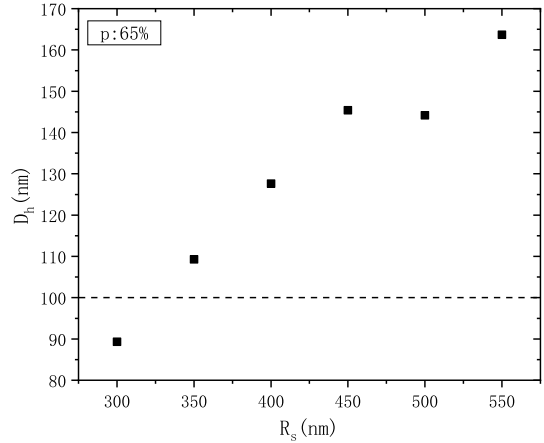}
  \caption{ \label{fig4}The packing fraction p=65$\%$. The simulated sintered particle radius $R_s$ is the average pore size of porous media formed under 300, 350, 400, 450, 500, and 550 $nm$. The dotted line represents the optimal pore size $D_h$=100 $nm$ formed by sintered metal powder in the experiment.} 
\end{figure}

In order to improve the performance of the heat exchanger, the contact area between the sinter and liquid helium should be large, namely, the particle size should be as small as possible. In contrast, the pore space formed should be large enough to ensure that $^3$He quasiparticle can pass smoothly. Therefore, the proper size of the nanoparticle should be selected. 
In Ref. \cite{stecher1994study}, the preferred average pore size formed by sintered powder is found to be 100 $nm$, making each pore much larger than the $^3$He quasiparticle whose diameter is around 3\AA \cite{kleinert1980can}. Therefore, it is reasonable to choose the average pore size of 100 $nm$ or the average throat size of 10 $nm$ as the criteria in our simulations. Fig. \ref{fig3} (a) and (b) show that both average throat size and pore size decreased with increasing packing fraction, respectively. One can see that a smaller radius of nanoparticles results in a smaller throat size and pore size.
In Ref. \cite{zaccone2022explicit}, the maximum packing fraction of hard spheres randomly packing is 65.896$\%$. Therefore, we calculated the change of the average pore size with the sphere radius when the packing fraction is 65$\%$, as shown in Fig. \ref{fig4}, the sphere radius of 30$\sim$35 $nm$ is the optimal sintering particle radius, which is just a case.

In the experiment, we know the mass of the sintering metal powder and the volume of the heat exchanger. When the mass density of sintered metal is known, it is not difficult to calculate the packing fraction. Then, using our simulation calculation method, we can get the optimal particle radius for sintering to improve the performance of the heat exchanger.

\section{Conclusions}
In conclusion, the performance of sintered metal heat exchanger is very important for realizing extremely ultra-low-temperature environments. The simulation method in this paper has guiding significance for the preparation of high-efficiency heat exchangers in experiments and  is beneficial to explore the method of reducing the Kapitza thermal resistance with liquid helium(pure $^3$He or $^3$He-$^4$He mixtures). We believe that the simulation results will be useful for our research work, and relevant studies are ongoing. One needs to pay attention to the fact that the heat exchanger formed by actual sintering is affected by sintering conditions and external operations, and its interior is more complicated. The influence of additional conditions has not been considered in the simulation method in this paper. Further research is needed in this direction.
\bibliography{ref.bib}

\end{document}